\documentclass{elsart}

\usepackage{graphicx}
\usepackage{amssymb}
\usepackage{multirow}
\begin{document}
\begin{frontmatter}

\title{Evaluation of cosmogenic production of $^{39}Ar$ and $^{42}Ar$ for rare-event physics using underground argon}
\author{C. Zhang}
\ead{Chao.Zhang@mountmarty.edu}
\address{Division of Natural Sciences, Mount Marty University, Yankton, SD 57078}
\author{D.-M. Mei},
\ead{Dongming.Mei@usd.edu}
\address{Department of Physics, The University of South Dakota, Vermillion, SD 57069}
\begin{abstract}
Underground argon (UAr) with lower cosmogenic activities of $^{39}Ar$ and $^{42}Ar$ has been planned as a detector in detecting 
scintillation light and charge collection using time projection chambers for dark matter searches and as a veto detector in suppressing backgrounds for neutrinoless double 
beta decay (0$\nu\beta\beta$) experiments. Long-lived radioactive isotopes, $^{39}Ar$ and $^{42}Ar$, can also be produced on the surface when 
UAr is pumped out from a deep well. Understanding the production of long-lived isotopes in Ar is 
important for utilizing UAr for dark matter and  0$\nu\beta\beta$ experiments in terms of 
its production, transportation, and storage. Ar exposure to cosmic rays at sea-level is simulated using Geant4 for a given 
cosmic ray muon, neutron, and proton energy spectrum.
We report the simulated cosmogenic production rates of $^{39}Ar$, $^{42}Ar$, and other long-lived isotopes at sea-level 
from fast neutrons, high energy muons, and high energy protons. Total production rates of 938.53/kg$_{Ar}\cdot$day and 5.81$\times$10$^{-3}$/kg$_{Ar}\cdot$day for $^{39}$Ar and $^{42}$Ar are found from our simulation. Utilizing these production rates, we set a time limit of 954 days constrained by the production of $^{39}$Ar for UAr to be on  the surface before it compromises the sensitivity for a dark matter experiment. Similarly, a time limit of 1702 days constrained by the production of $^{42}$Ar is found for a 0$\nu\beta\beta$ experiment. 

\end{abstract}


\end{frontmatter}
% main text
\section{Introduction}
Observational evidence indicates that dark matter accounts for approximately 85\% of the matter in our 
universe\cite{blumenthal, davis, bennett}. 
But the nature of dark matter is still mysterious. Of all the dark matter candidates, the 
Weakly Interacting Massive Particle (WIMP) seems promising and is favored by scientific communities. 
Assuming the WIMP is a new elementary particle, which is ubiquitous in galaxies, it could be directly detected through its 
elastic recoils on ordinary matter\cite{goodman, feng}. 
Many dark matter experiments are actively seeking WIMP scattering events but none have succeeded so far~\cite{cdms, cdex, cogent, cresst, coupp, damic, darkside, drift, edelweiss, kim, lux, pandax, pico, supercdms, xenon10, xenon100, xenon1t, xmass, zeplin, deap}. DAMA/LIBRA has been claiming for decades that the observation of the annual modulation in the detection is the recoil signal stroked by WIMPs~\cite{dama}, although their claim is not confirmed by any other experiments.
To further push dark matter detection sensitivities, the next generation large-scale dark matter experiments are being built aiming at the WIMP-nucleon cross section down to the level of $ 10^{-48} \, \rm{cm}^{2}$~\cite{lz, xenonnt, darkside20k}.  This level of 
sensitivity also reaches the floor of the neutrino background due to coherent neutrino induced nuclear recoils in the detector~\cite{lgs, agu}.    In order to achieve such detection sensitivity, large-scale detectors with hundreds of ton-years exposure under an ultra-low background environment are needed. For example, DarkSide-20k, which will use underground argon (UAr) as the target~\cite{darkside20k}, is a two-phase liquid Ar detector with a 23 tonnes active volume.      

\par
The experiments seeking to measure the half-life of neutrinoless double-beta decay ($0\nu\beta\beta$) aim to determine the Majorana nature of the neutrino and 
help understand the absolute neutrino masses and their mass hierarchy\cite{steve1, steve2, avignone, barabash}. 
No discovery has been made by current $0\nu\beta\beta$ experiments with the sensitivity of the decay half-life 
up to $\sim 10^{26}$ years~\cite{gerda}. This sets an upper limit on the effective Majorana mass of the
electron neutrino less than 0.2 eV\cite{gerda, exo200, kamland, sno+, majorana}. A future large-scale experiment, LEGEND-1000, with increased sensitivity 
up to $10^{28}$ years has been proposed~\cite{legend}. This aims for a sensitivity of 0.01 event per ton per year in the region
of interest. Aiming at such detection sensitivity, instrumental backgrounds from radioactive isotopes 
need to be well suppressed and understood. LEGEND-1000 plans to use UAr as the active veto detector. 

\par 
Both dark matter and $0\nu\beta\beta$ experiments require their background rate at the 
region of interest to be extremely low. The instrumental background from long-lived radioactive isotopes in detector components
must be minimized and accurately measured. In addition, cosmogenic activation can add more radioactivities 
to the background budget. 
Liquid Ar, a relatively cheap noble liquid scintillator, is a widely used medium for the detection of ionizing radiation.
Taking advantage of its high ionization yield and characteristic scintillation time profile~\cite{mark}, liquid Ar is utilized as an active target
or scintillation veto material in dark matter searches and $0\nu\beta\beta$ experiments~\cite{darkside, deap, darkside20k, gerda, legend}. The problematic issue is that there are long-lived radioactive isotopes existing in Ar when it is produced on the surface. Both $^{39}$Ar (t$_{1/2}$=269 years) and $^{42}$Ar (t$_{1/2}$=32.9 years) are produced through cosmogenic activation~\cite{asb1, asb2}. The isotopic abundance of $^{39}$Ar in atmospheric argon is at a level of (8.1$\pm$0.6) $\times$10$^{-16}$g/g, which results in a decay rate of $\sim$1 Bq/kg in a liquid Ar detector~\cite{darkside, deap}. It is a pure $\beta$ decay to a stable daughter nucleus ($^{39}$K). This decay rate and the $\beta$ decay Q value of 565.5 keV produce background and pile-up concerns in the detectors for dark matter searches. On the other hand, $^{42}Ar$ was measured by the G{\sc{erda}} collaboration and the isotopic abundance is determined at a level of (9.2$^{+2.2}_{-4.6}$)$\times$10$^{-21}$ g/g. More recent measurement from the DEAP collaboration shows the specific activity of $^{42}$Ar to be 40.4 $\pm$ 5.9 $\mu$Bq/kg~\cite{deep42ar} which is lower than the GERDA’s result of (88$^{+21}_{-44})$ $\mu$Bq/kg. Its decay product, $^{42}$K, a $\beta$ emitter with a decay Q value of 3.52 MeV, is a background to 0$\nu\beta\beta$ decay experiments when using liquid argon as a veto detector~\cite{gerda}. 
 Note that a fraction of $^{42}$Ar could be produced in the Earth’s atmosphere due to nuclear bomb tests from 1945 to 1962~\cite{asb1}. The concentration of $^{42}$Ar after bomb testing is estimated to be $\sim$ 10$^{-22}$ to $\sim$10$^{-23}$ atoms per $^{nat}$Ar atoms~\cite{asb1}. This is only $\sim$1\% to $\sim$10\% of the measured concentration of $^{42}$Ar~\cite{deep42ar, asb1}. The current $^{42}$Ar in Earth’s atmosphere is dominated by spallation reactions from the cosmic ray nucleon component through $^{40}$Ar($\alpha, 2p)^{42}$Ar process.

To obtain Ar with a lower level of $^{39}$Ar and $^{42}$Ar, it is natural to explore Ar which has been deep underground where Ar has existed for thousands of years~\cite{mei}. Since the production of $^{39}$Ar depends on the depth of underground wells and the surrounding rock condition in terms of its porosity, UAr may not always have a low level of $^{39}$Ar~\cite{mei}. Because the production of $^{42}$Ar is mainly through $^{40}$Ar($\alpha$,2p)$^{42}$Ar and the interaction threshold energy of $\alpha$ particles is above 12 MeV, it is expected that the production of $^{42}$Ar is strongly suppressed in deep underground since there are no $\alpha$ particles with energy greater than 12 MeV from natural radioactivity decays. Amazingly, the UAr with low radioactivity produced by the DarkSide collaboration~\cite{darkside50} in Colorado of the United States has shown a reduction of 1400 in $^{39}$Ar when compared to atmospheric Ar. Note that the UAr produced elsewhere may not have such a low level of $^{39}$Ar~\cite{mei}. Although there is no measured reduction factor of $^{42}$Ar reported from the UAr produced by DarkSide, it is expected that the reduction of $^{42}$Ar should be at least a factor of 1400~\cite{legend}. Therefore, the UAr produced by DarkSide in Colorado is particularly valuable for DarkSide-20k and LEGEND-1000. To keep low level of $^{39}$Ar and $^{42}$Ar for underground experiments, any potential production processes should be carefully examined and suppressed since $^{39}$Ar and $^{42}$Ar are mainly produced through cosmogenic processes when Ar is on the surface~\cite{asb1, asb2}, 
precise calculations of cosmogenic activation are needed to determine the maximum tolerable surface exposure time
of argon during its production, transport and storage.
\par
In this paper, we evaluate cosmogenic production of radioactive isotopes in Ar. 
The production rates of $^{39}Ar$, $^{42}Ar$, and other long-lived isotopes
from fast neutrons, high energy muons, and high energy protons are obtained using Geant4-based simulations.
The results are also compared with some experimental data.  

\section{Evaluation of cosmogenic production in Ar}

\subsection{Evaluation tools and input sources}
GEANT4 (V10.7p02) with shielding modular physics list \cite{geant4, shielding} is used for this study.
It includes
a set of electromagnetic and hadronic physics processes for 
high energy or high precision simulation needs. 
G4MuonNuclearProcess is activated to simulate the muon-nuclear interactions.  
\par
\begin{figure}
\includegraphics[width=0.95\textwidth]{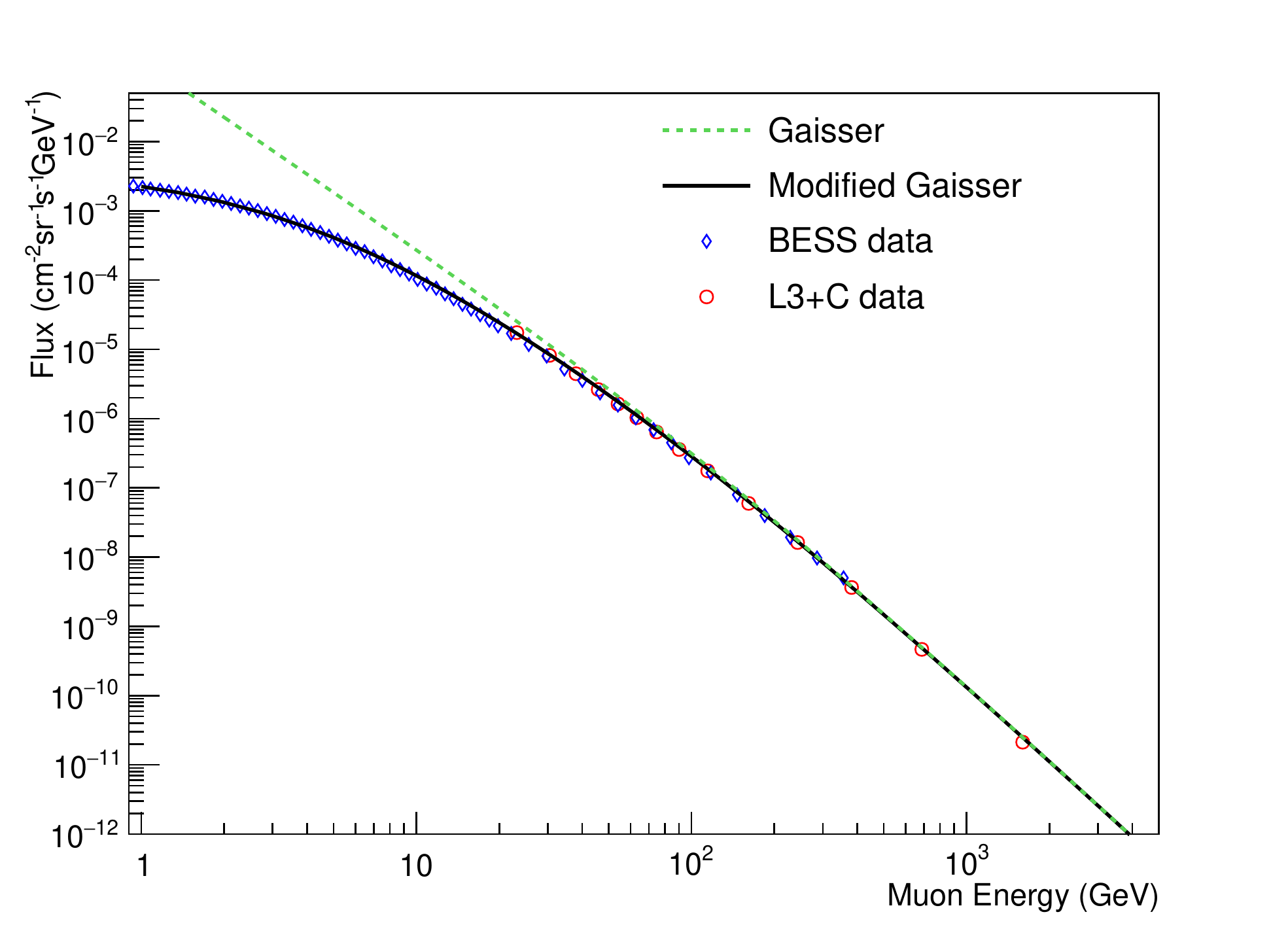}
\caption{\label{muon}
Vertical muon flux at sea level. The dashed line is given by Gaisser's formula\cite{gaisser}. The solid line stands for 
the modified Gaisser's formula\cite{guan}. As a comparison, the measured muon distribution at $0^{\circ}$ 
zenith angle from L3+C\cite{l3c} and BESS\cite{bess} are also presented. 
}
\end{figure}

Cosmic ray muon flux at sea level can be parameterized using Gaisser's formula\cite{gaisser}. 
It assumes a flat Earth which is only valid for muons with energy $E_{\mu}>100 GeV$ (see Fig. \ref{muon}).   
In our simulation, the input muons are sampled by using the modified Gaisser's formula\cite{guan}.
The total muon flux is normalized to 0.015 $cm^{-2}s^{-1}$ with energy spanning from 1 GeV to 100 TeV. 
Stop muons are not included in the simulation due to their relatively small contribution to the 
total cosmogenic activation budget for Ar. 
\par
The input fast neutrons with energy greater than 4 MeV are 
adopted from the ground-based neutron flux measurements at a reference location at sea level,  
New York City (NY data)\cite{gordon}. 
The total flux is normalized to be 0.004 $cm^{-2}s^{-1}$. It is worth mentioning that
the MeV neutrons may slightly vary from location to location due to local radioactivities
such as neutron yield from $(\alpha, n)$ process in materials.   
 
\par
\begin{figure}
\includegraphics[width=0.95\textwidth]{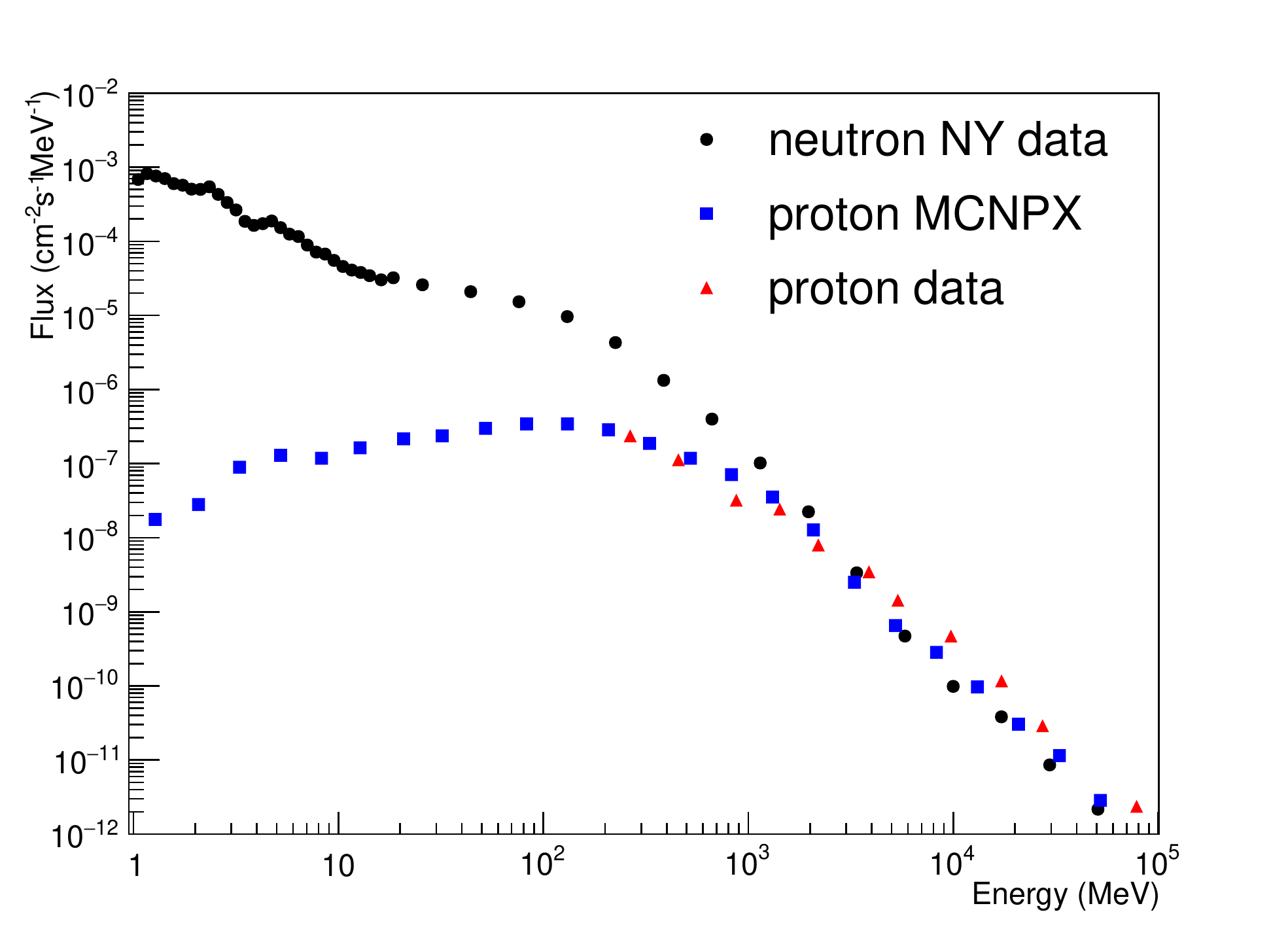}
\caption{\label{neuProtonComp}
Comparison of neutron and proton energy distribution at sea level. Black dots are experimental neutron data measured at a reference location
at New York City (NY data)\cite{gordon}. Blue squares are Monte Carlo generated proton spectrum using MCNPX code\cite{chrisLLNL} with its flux
normalized to measured proton data (red triangles\cite{brooke}). 
}
\end{figure}

Cosmic ray proton induced activation in Ar is also simulated although
the number of protons below a few GeV is much less than the number of neutrons in the atmosphere (see Fig. \ref{neuProtonComp}).  
The proton spectrum is harder than the neutron one with energies exceeding a few GeV. 
The input cosmic ray proton flux is adopted from the result of MCNPX \cite{chrisLLNL} simulation code. Protons with energy $E_{p}> 10$ MeV 
are sampled for our use.   

\subsection{Cosmogenic activation in Ar}
Commercial Ar is mostly extracted from the Earth's atmosphere. It consists primarily of the 
stable isotopes $^{40}Ar$, $^{36}Ar$, and $^{38}Ar$.  
Although natural argon is stable, it suffers from cosmic-ray bombardment at the earth's surface.  
Ar radionuclides produced by cosmogenic activation, such as
$^{37}Ar$, $^{39}Ar$ and $^{42}Ar$, are irreducible and turn out to be 
 a significant source of background for low-background Ar-based radiation detectors.  
To effectively suppress cosmogenic activation, switching from atmospheric Ar to deep UAr had been proposed for DarkSide-20k and LEGEND-1000. 
Significant production of ultra-low background Ar extracted and purified from deep $CO_{2}$ wells has been successfully developed. It achieves a factor of 1400 reduction (or 0.73 $\rm{mBq/kg_{Ar}}$) over 
atmospheric argon in terms of their $^{39}Ar$ concentration\cite{darkside2016}.  
However, the cosmogenic production rates are still critical and need to be accurately calculated to optimize 
the surface exposure time during its production, shipping and storage. Note that $^{37}$Ar is not so dangerous for dark matter and 0$\nu\beta\beta$ decay experiments due to its shorter half-life (35 days).

\par
A comprehensive Geant4 (V10.7p02) simulation has been conducted to evaluate the cosmogenic activation rate in an Ar target. 
Geant4 default material "G4\_lAr" ($^{40}$Ar:99.604\%; $^{36}$Ar:0.334\%; $^{38}$Ar:0.063\%) is selected and stored in a thin cylinder tank that is 2.0 m in diameter and 2.0 m in height.
The liquid argon has the density of 1.40 $g/cm^{3}$ which gives the total mass of $\sim$8.80 ton per cylinder. 
The cosmic-ray muons, neutrons and protons are considered as inputs at the top of the Ar target. 

\par
The production of $^{39}$Ar in $^{40}$Ar is through the reaction of $^{40}$Ar(n, 2n$^{'}$)$^{39}$Ar when high energy neutrons ($>$10 MeV) bombard the Ar target. This reaction is expected to be seen in the Geant4 simulation. A main purpose of this simulation is to determine if $^{42}$Ar can be seen in the Geant4 simulation since the production rate of $^{42}$Ar is extremely small. The long-lived argon isotope $^{42}Ar$ undergoes a $\beta$ decay with a half-life of 32.9 years. The $\beta$ decay of its daughter isotope, $^{42}K$, has a maximum electron energy of 3.52 MeV
 which becomes a vital background for $0\nu\beta\beta$ experiments in the region of interest ($Q_{\beta\beta}=2.039 MeV$ for $0\nu\beta\beta$ decay from $^{76}$Ge)\cite{legend} when using liquid Ar as a veto detector.  It is understood that
$^{42}Ar$ is mainly produced through the reaction channel $^{40}Ar(\alpha, 2p)^{42}Ar$ with a Q-value of -12.77 MeV\cite{asb1, asb2}. This means that the $\alpha$ particles that can generate $^{42}$Ar in $^{40}$Ar must have kinetic energies greater than 12.77 MeV. 

In order to see the production of $^{42}$Ar in our simulation in a reasonable computing time (one month) in the high precision computing cluster with 72 CPUs at the University of South Dakota, all particles with energy less than 12 MeV are killed (tracks stop and kill). Therefore, our results exclude all capture processes. For example $^{36}$Ar(n,$\gamma$)$^{37}$Ar is not found in our simulation. We found that the cosmogenic production rate of $^{42}Ar$ results in a rate of  $5.81\times 10^{-3} \, \rm{atoms/kg_{Ar}/day}$ in total which  corresponds to the activity of 
$3.88\times 10^{-12} \, \rm{Bq/kg_{Ar}/day}$. This shows that the Geant4 simulation is able to predict the production of $^{42}$Ar. We summarize the three long-lived argon isotopes in Table 1 and other long-lived isotopes in Table 2. 
%%%%%argon radionuclides
\begin{table}[htbp]
\caption{\label{Argon}
Cosmogenic activation rates of three argon isotopes: $^{37}Ar$, $^{39}Ar$ and $^{42}Ar$. The simulation results are also compared with
the ones from measurements and estimations\cite{ArgonData}. } 
%\centering
\begin{tabular}{l|ccc}\hline
          & $^{37}Ar$  & $^{39}Ar$ & $^{42}Ar$  \\
	  & \multicolumn{3}{c}{$\rm{atoms/kg_{Ar}/day}$} \\
\hline
Neutrons (this work)  &176.01 & 857.73 & $4.60\times 10^{-3}$       \\
Neutrons (measurement\cite{ArgonData}) & $51.0\pm 7.4$ & $759\pm128$ & -       \\
\hline
Muons (this work)    & 2.40 & 52.27 & $1.57\times 10^{-4}$       \\
Muons (calculation\cite{ArgonData})    & - & $172\pm 26$ & -       \\
\hline
Protons (this work) &6.20 & 28.53 &$1.05\times 10^{-3}$       \\
Protons (calculation\cite{ArgonData}) & $1.73\pm0.35$& $3.6\pm 2.2$  & -       \\
\hline
Total (this work) &184.61 & 938.53  &$5.81\times 10^{-3}$       \\
Total (Ref.\cite{ArgonData}) & $52.73\pm7.75$& $934.6\pm 156.2$  & -       \\
\hline
\end{tabular}
\end{table}

%%%%%Other radionuclides
\begin{table}[htbp]
\caption{\label{Others}
Cosmogenic activation rates of other long-lived isotopes. }
%\centering
\begin{tabular}{l|ccc}\hline
 Isotope, Half Life & Neutron & Muon & Proton   \\
		    & \multicolumn{3}{c}{$\rm{atoms/kg_{Ar}/day}$ } \\
\hline
$^{3}H$, 12.32y & $3.00\times 10^{-4}$ & $6.56\times 10^{-6}$ & $1.05\times 10^{-4}$  \\
$^{7}Be, 1.387\times 10^{6}y $ & $3.43\times 10^{-3}$ & $6.47\times 10^{-3}$ & $9.62\times 10^{-3}$  \\
$^{10}Be, 53.22d $ & $7.05\times 10^{-3}$ & $5.22\times 10^{-3}$& $1.10\times 10^{-2}$ \\
$^{14}C, 5.703\times 10^{3}y $ & 0.10& $9.85\times 10^{-3}$& $4.56\times 10^{-2}$ \\
$^{22}Na, 2.602y $ & 0.37& $2.35\times 10^{-2}$& $9.92\times 10^{-2}$ \\
$^{26}Al, 7.17\times 10^{5}y $ & 0.63& $4.24\times 10^{-2}$& 0.12 \\
$^{32}Si, 153y $ & 7.00& 0.14& 0.47  \\
$^{32}P, 14.268d $ & 15.7& 0.38&1.05  \\
$^{33}P, 25.3d $ & 22.5& 0.42& 1.32 \\
$^{35}S, 87.37d $ & 74.5& 1.66&3.18  \\
$^{36}Cl, 3.01\times 10^{5}y $ & 75.5& 1.23&3.32  \\
$^{40}K, 1.248\times 10^{9}y $ & 1.80& $5.86\times 10^{-2}$& 0.56 \\
$^{41}Ca, 9.94\times 10^{4}y $ & $1.85\times 10^{-3}$& $1.02\times 10^{-4}$& $5.68\times 10^{-4}$ \\
\hline
\end{tabular}
\end{table}

As can be seen from the results shown in Table 1 and 2, the production rates of long-lived isotopes in Ar is dictated by high energy neutrons. The production rate of $^{39}$Ar by fast neutrons from this work is in a reasonable agreement with the measurement made by Saldanha et al.~\cite{ArgonData}. However, the production rate of $^{37}$Ar by neutrons is different from Saldanha et al. by a factor of $\sim$3. The discrepancy is caused by the different $^{40}$Ar(n,4n)$^{37}$Ar production cross section used in the process. This interaction cross section has not been experimentally measured. 

\begin{figure}
\includegraphics[width=0.95\textwidth]{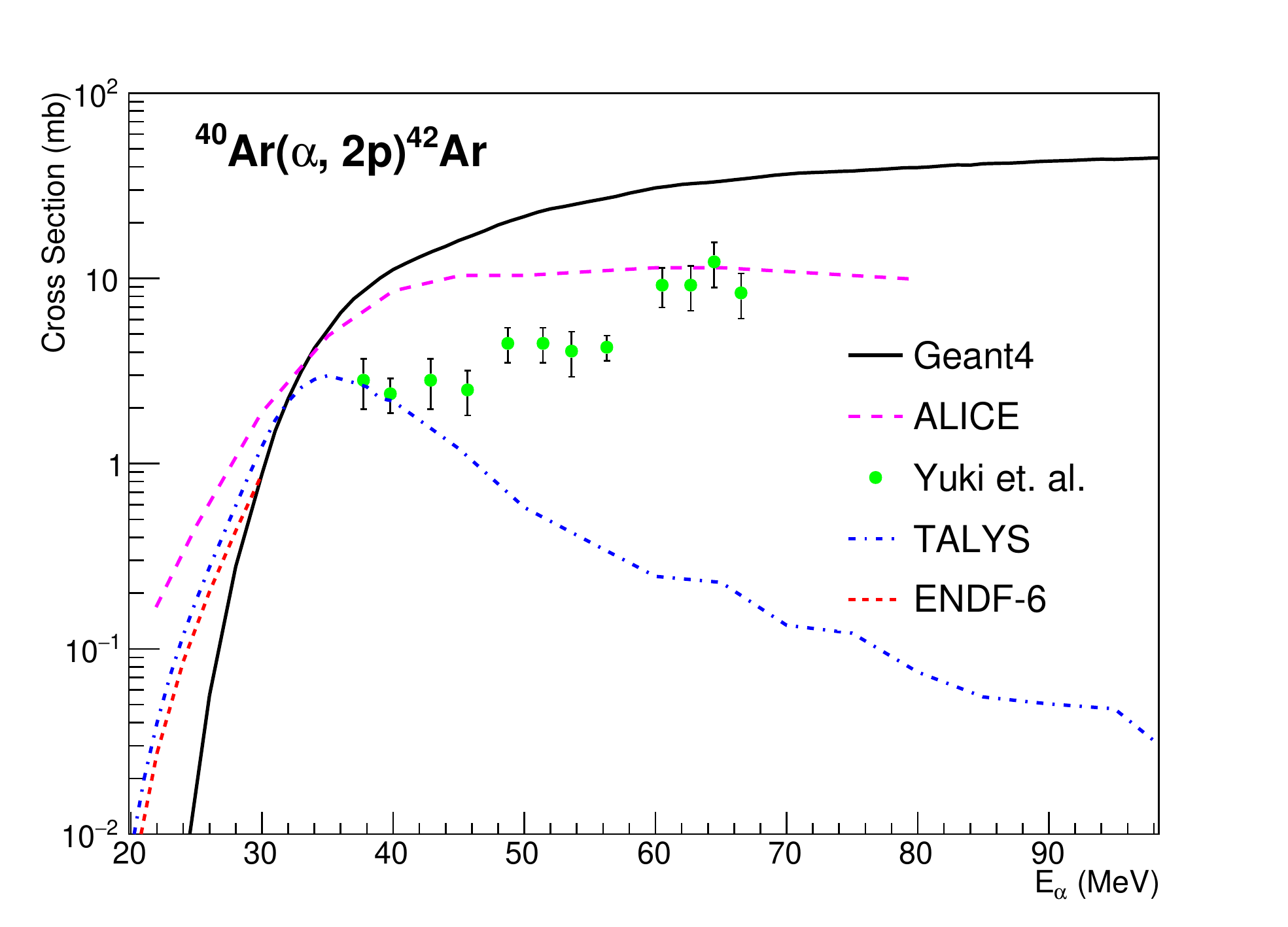}
\caption{\label{Ar42xSectionComp}
Shown is a comparison for the cross section of $^{40}\rm{Ar}(\alpha, 2p)^{42}\rm{Ar}$ extracted from Geant4 with the experimental measurement (Yuki et al.~\cite{yuki}), 
as well as that of ALICE~\cite{alice}, TALYS~\cite{talys} and ENDF-6~\cite{endf}.   
}
\end{figure}

The total production rate of $^{39}$Ar measured by Saldanha et al.~\cite{ArgonData} (934.6$\pm$156.2 atoms/kg$\cdot$day) is in a good agreement with this work (938.53 atoms/kg$\cdot$day) using the Geant4 simulation. To validate the simulated production rate of $^{42}$Ar from this work, we compared the cross section of  $^{40}\rm{Ar}(\alpha, 2p)^{42}\rm{Ar}$ from a measurement made by Yuki et al.~\cite{yuki}, the ALICE code~\cite{alice}, a nuclear reaction simulator - TALYS~\cite{talys}, and a nuclear database - ENDF~\cite{endf} as shown in Figure~\ref{Ar42xSectionComp}. The cross section extracted from Geant4 increases as the kinetic energy of $\alpha$ particles increases. This tendency is similar to the experimental data~\cite{yuki} and the ALICE code. However, the value of the cross section extracted from Geant4 is different by a factor of 2 to 5 (depending on energy) from the available data provided by Yuki et al. and a factor of $\sim$2 from the ALICE code. Note that the cross section obtained from the TALYS code is significantly different from Geant4, the ALICE code, and the available data. The ENDF-6 database only provides the cross section for energy below 30 MeV. Thus, we can conclude that the cross section of $^{40}\rm{Ar}(\alpha, 2p)^{42}\rm{Ar}$  is not well understood and more experimental data are needed. 
 Nevertheless, the Geant4 simulation confirms the production of $^{42}$Ar on the Earth's surface through the secondary $\alpha$ particles from cosmic ray muons, neutrons, and protons. Note that the calculated rates of $^{42}$Ar production in the atmospheric argon at sea level cannot be directly compared to the experimentally observed abundance of $^{42}$Ar. This is because $^{42}$Ar is mainly produced at the top of the atmosphere where the primary cosmic ray fluxes are significantly larger than terrestrial cosmic ray fluxes~\cite{ziegler}. In addition, the energy spectra of the primary cosmic ray fluxes at the top of the atmosphere are much harder than that of the cosmic rays at sea level. Therefore, the experimental observed abundance of $^{42}$Ar is significant larger than the calculated rates of $^{42}$Ar production at sea level.
 
\subsection{Tolerable Exposure Time Limit}

One can use the production rates of $^{39}$Ar and $^{42}$Ar obtained from this work to set the allowed exposure time for UAr on the surface in terms of production, transportation, and storage. Since the UAr produced by DarkSide in Colorado is depleted in $^{39}$Ar by a factor of 1400 when compared to atmospheric Ar, we can also expect that the concentration of $^{42}$Ar is depleted by at least a factor of 1400. Therefore, the cosmogenic production of $^{39}$Ar and $^{42}$Ar in the UAr should be controlled in a level of less than 10\% of the existing concentration of $^{39}$Ar and $^{42}$Ar in the UAr produced by DarkSide. This level of additional cosmogenic production in the UAr will not compromise its sensitivity for the planned DarkSide-20k and LEGEND-1000. 

Using a constraint of 10\% additional cosmogenic production of $^{39}$Ar and $^{42}$Ar in the UAr, we estimate the allowed exposure time using the formula below~\cite{ken}: 
\begin{equation}
 \label{eq1}
    N(t) = \frac{R}{\lambda}[1-exp(-\lambda t)],
\end{equation}
where $N(t)$ is the additional number of atoms (10\% of the existing $^{39}$Ar or $^{42}$Ar in the UAr) produced in the UAr through cosmogenic activation when the UAr is on the surface during the production, transportation, and storage, $R$ is the production rate from the Geant4 simulation, $\lambda$ is the decay constant, and $t$ is the allowed exposure time. Note the 10\% of the existing $^{39}$Ar in the UAr is calculated using the atmospheric level of 8.1$\times$10$^{-16}$g/g divided by 1400 and multiplied by 10\%. Similarly, the 10\% of the existing $^{42}$Ar in the UAr is calculated using the atmospheric level of 9.2$\times$10$^{-21}$g/g divided by 1400 (assuming an upper limit for a conservative consideration) and multiplied by 10\%. 

Derived from Eq~(\ref{eq1}), the allowed exposure time can be expressed as:
\begin{equation}
\label{eq2}
    t = \frac{ln[1-N(t)\lambda/R]}{-\lambda}.
\end{equation}
Using Eq~(\ref{eq2}), the allowed exposure time for the UAr on the surface are 954 days in terms of the production of $^{39}$Ar and 1702 days in terms of the production $^{42}$Ar, respectively. This means that the UAr can be on the surface for more than two years, which is a quite reasonable time to produce enough UAr before transporting to underground for storage. Note that the cosmic ray neutron flux on the Earth’s surface varies with altitude~\cite{ziegler}, which will affect the production of $^{39}$Ar and $^{42}$Ar accordingly. The exposure time limit will change with respect to different altitudes.

\begin{figure}
\includegraphics[width=0.95\textwidth]{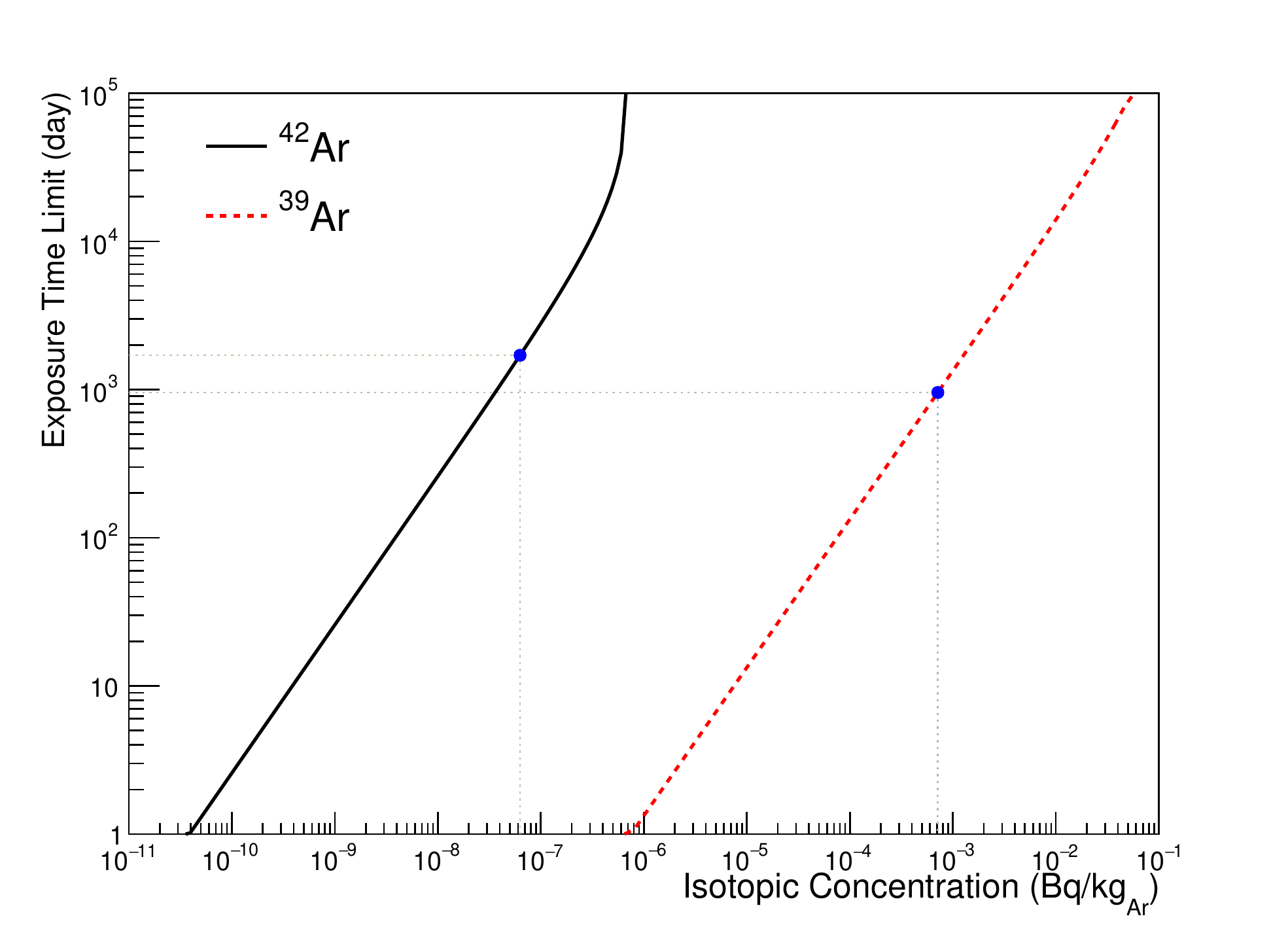}
\caption{\label{exposureTime}
Exposure time limit versus depleted $^{42}Ar/^{39}Ar$ concentration. The blue dots represent the reduction factor of 1400 in terms of its $^{42}Ar$ and $^{39}Ar$ concentration in atmospheric argon, respectively.
}
\end{figure}

 One can also show the calculated exposure limits using Eq.~\ref{eq2}, for both $^{42}$Ar and $^{39}$Ar production at sea level, versus a target activity level since different experiments will have their own target activity levels. Figure~\ref{exposureTime} displays the results.

\section{Summary}
The cosmogenic activations of Ar for the next generation rare event search 
experiments at sea level have been simulated using 
GEANT4 package. Fast neutrons, muons and protons at the earth surface 
are considered individually. 
The activation rates of $^{37}Ar$, $^{39}$Ar and $^{42}Ar$ are compared 
with the measurement and model predictions\cite{ArgonData}. We found that the production rates of long-lived isotopes through cosmogenic activation on the earth's surface are dictated by fast neutrons. Using the production rates of $^{39}$Ar and $^{42}$Ar from this work, we set a limit of the exposure time for the UAr on the surface during its production, transportation, and storage. The limit is 954 days if the production of $^{39}$Ar is a main driver for a dark matter experiment such as DarkSide-20k. On the other hand, the limit is 1702 days if the production of $^{42}$Ar is a main concern for a 0$\nu\beta\beta$ decay experiment such as LEGEND-100 using UAr as a veto. We conclude that the UAr from Colorado discovered by DarkSide can be exposed on the earth's surface in terms of production, transportation, and storage for more than 900 days without compromising the sensitivity of planned DarkSide-20k and LEGEND-1000 experiments.  

\section{Acknowledgement}
The authors would like to thank Dr. Christina Keller for a careful reading of this manuscript. This work is supported by NSF OISE-1743790, PHYS-1902577, OIA-1738695, 
DOE FG02-10ER46709, the Office of Research at the University of South Dakota and a research center supported by the State of South Dakota.

\end{document}